\documentclass[12pt,twoside]{article}
\usepackage[mathscr]{eucal}
\usepackage{amsmath,amsfonts,amssymb,amsthm,mathabx,empheq}
\usepackage{times}
\usepackage{pdfsync}
\usepackage{cite}
\usepackage{url}
\usepackage{hyperref}
\usepackage{tensor}
\usepackage{color}
\usepackage{multicol}
\usepackage{bbold}

\advance\textheight\topskip
\allowdisplaybreaks[1]

\newcommand\td{\text{d}}

\newcommand{\p}{\partial}

\newcommand{\be}{\begin{equation}}
\newcommand{\ee}{\end{equation}}
\newcommand{\bea}{\begin{eqnarray}}
\newcommand{\eea}{\end{eqnarray}}

\def\bz{\bar z}

\def\ga{\gamma_{z\bz}}

\def\cH{{\cal H}}

\def\n{\nabla}

\newcommand*\xbar[1]{%
  \hbox{%
    \vbox{%
      \hrule height 0.5pt 
      \kern0.3ex
      \hbox{%
        \kern-0.0em
        \ensuremath{#1}%
        \kern-0.0em
      }%
    }%
  }%
}

\hypersetup{
colorlinks=true,
linktoc=page,    
linkcolor=blue,
citecolor=blue,
urlcolor=blue,
colorlinks=true,
}

\newcommand\RR{\ensuremath{\mathbb R}}

\DeclareFontFamily{OT1}{rsfs}{} \DeclareFontShape{OT1}{rsfs}{m}{n}{
<-7> rsfs5 <7-10> rsfs7 <10-> rsfs10}{}
\DeclareMathAlphabet{\mycal}{OT1}{rsfs}{m}{n}

\begin{document}
\title{Near horizon linearized gravitaty and soft theorem}

\author{Pujian Mao, Kai-Yu Zhang, Bochen Zhou}

\date{}

\def\mytitle{Near horizon linearized gravity and soft theorem}

\addtolength{\headsep}{4pt}

\begin{centering}

  \vspace{1cm}

  \textbf{\Large{\mytitle}}

  \vspace{1.5cm}

  {\large Pujian Mao,  Kai-Yu Zhang, Bochen Zhou}

\vspace{.5cm}

\vspace{.5cm}
\begin{minipage}{.9\textwidth}\small \it  \begin{center}
    Center for Joint Quantum Studies and Department of Physics,\\
     School of Science, Tianjin University, 135 Yaguan Road, Tianjin 300350, China
 \end{center}
\end{minipage}

\end{centering}


\vspace{1cm}

\begin{center}
\begin{minipage}{.9\textwidth}
  \textsc{Abstract}. In this paper, we study the linearized gravity theory in the near horizon region of the  Schwarzschild black hole in four dimensional spacetime. Under the Newman-Unti gauge, we derive the most general near horizon symmetry and solution space without any near horizon fall-off condition. There are four towers of surface charges that are generic functions on the horizon associated to the near horizon symmetry. With suitable near horizon fall-off conditions, we reveal a soft graviton theorem from the Ward identity of the near horizon supertranslation in both coordinates space and momentum space. 
\end{minipage}
\end{center}

\begin{center}
Emails: pjmao@tju.edu.cn,\, kaiyu\_zhang@tju.edu.cn,\,zhoubch@tju.edu.cn
\end{center}

\thispagestyle{empty}

\newpage
\tableofcontents

\section{Introduction}

The existence of horizon is a very fascinating feature of geometries with Lorentzian signature. A stationary black hole with a future event horizon admits four mechanical laws which are very similar to the ordinary laws of thermodynamics \cite{Bardeen:1973gs}. Then, the discovery of the Hawking radiation \cite{Hawking:1975vcx} makes the laws of black hole physics real thermodynamical laws. Hence, the process of black hole formation equivalently is a process of thermalization. From a quantum perspective, the process of black hole formation starts from pure states and ends with a thermal state, which means that the evolution is not unitary. This invokes the black hole information paradox \cite{Hawking:1976ra}. It is a long-standing problem in black hole physics and quantum gravity. Recently, to resolve the black hole information paradox, Hawking, Perry, and Strominger (HPS) proposed that black holes can carry soft hairs \cite{Hawking:2016msc}. The presence of soft hairs are required from the charge conservation of asymptotic symmetries which triggers quantum correlations between the outgoing Hawking quanta and the soft hair configuration. 

In the original proposal of HPS, it is assumed that the soft hairs are soft gravitons or photons on the black hole horizon. Such observation is based on a fascinating triangular equivalence \cite{Strominger:2017zoo}, where soft theorems are nothing but the quantum Ward identity of asymptotic symmetries \cite{Strominger:2013lka,Strominger:2013jfa,He:2014laa,He:2014cra,He:2015zea}. Hence, an asymptotic symmetry transformation on black hole spacetime is equivalent to implanting soft hairs on the black hole \cite{Hawking:2016msc,Compere:2016jwb,Compere:2016hzt,Mao:2016pwq,Afshar:2016uax,Mirbabayi:2016axw,Grumiller:2016kcp,Donnay:2016ejv,Cai:2016idg,Hawking:2016sgy,Afshar:2016kjj,Choi:2018oel,Choi:2019fuq}. However, the soft theorems in the triangular relation are in flat spacetime. The corrections from the presence of a black hole was not considered in the original HPS proposal. Moreover, a black hole admits an event horizon which can be considered as an inner boundary of the spacetime. So the connections between symmetries at null infinity and soft theorems in flat spacetime may not be complete for the black hole soft hair proposal. One ongoing direction to fill in those gaps is the study of soft theorems in curved spacetime derived from the near horizon symmetry \cite{Cheng:2022xyr,Cheng:2022xgm,Mao:2023rca}. In particular, it is proven that the soft electric hairs of the Schwarzschild spacetime are indeed implanted by soft photons on the horizon which are derived from the Ward identity of the near horizon large gauge transformation \cite{Cheng:2022xyr}. 
In this work, we will extend the interplay between near horizon symmetry and soft theorems in \cite{Cheng:2022xyr,Cheng:2022xgm} to linearized gravity theory. The reason of linearisation is twofold. On the one hand, the interplay that we are interested is proposed for resolving the black hole information paradox. So we need a fixed black hole background. On the other hand, one needs to involve mode expansions to derive a soft theorem which is completely unclear how to perform for a gravitational theory with the principle of equivalence. Hence, the obvious choice is to study linearized gravity theory in a black hole spacetime.  

In this paper, we study the linearized gravity theory in the near horizon region of the Schwarzschild spacetime in four dimensions. The Newman-Unit (NU) gauge \cite{Newman:1962cia} is adapted to the linearized theory. We derive the residual gauge transformation that preserves the NU gauge conditions without imposing any fall-off condition. The most general residual (near horizon) symmetry consists of four arbitrary functions on the horizon which form a closed algebra for the linearized theory. Then we compute the solution space of the linearized theory and the surface charges that are associated to the near horizon symmetries. The connection of asymptotic symmetry and soft theorem is subject to proper fall-off conditions that the gauge or gravitational fields should satisfy \cite{Strominger:2013lka,Strominger:2013jfa,He:2014laa,He:2014cra,He:2015zea,Strominger:2017zoo}. This is a very subtle point in the near horizon region. Because there is no clear criterion to govern the near horizon behavior of the physical fields. A crucial fact that such connection can be successfully extended to the near horizon region for gauge theory in \cite{Cheng:2022xyr,Cheng:2022xgm} is that the fall-off conditions at null infinity can be easily adapted to the near horizon region in the sense that the near horizon solution space and the formula of the near horizon surface charges are very similar to the null infinity case. The situation is significantly changed for gravity. It is not obvious which type of near horizon conditions should be imposed such that the null infinity relations can be recovered from the near horizon analysis for the case of Einstein gravity \cite{Adami:2021nnf}. Such difficulty is inherited by the linearized theory. Remarkably, it is shown that there is a proper set of near horizon fall-off conditions such that it preserves the near horizon supertranslation \cite{Donnay:2015abr}, and the supertranslation charge can naturally split into the soft part and the hard part which will lead to a soft graviton theorem following the null infinity scenario \cite{He:2014laa} with adaptions to the near horizon computations \cite{Cheng:2022xyr,Cheng:2022xgm,Mao:2023rca}.    

The organization of this paper is very simple. In the next section, we derive the near horizon symmetry, solution space, and surface charges with respect to the NU gauge. In Section \ref{soft}, we introduce proper near horizon fall-off conditions. A near horizon soft graviton theorem is derived from the Ward identity of the near horizon supertranslation in the reduced solution space. We conclude in the last section. There is one Appendix where we present the details of the modified stress tensor.

\section{Near horizon linearized theory}
\label{solution}

\begin{figure}[ht] 
\center{\includegraphics[width=0.9\linewidth]{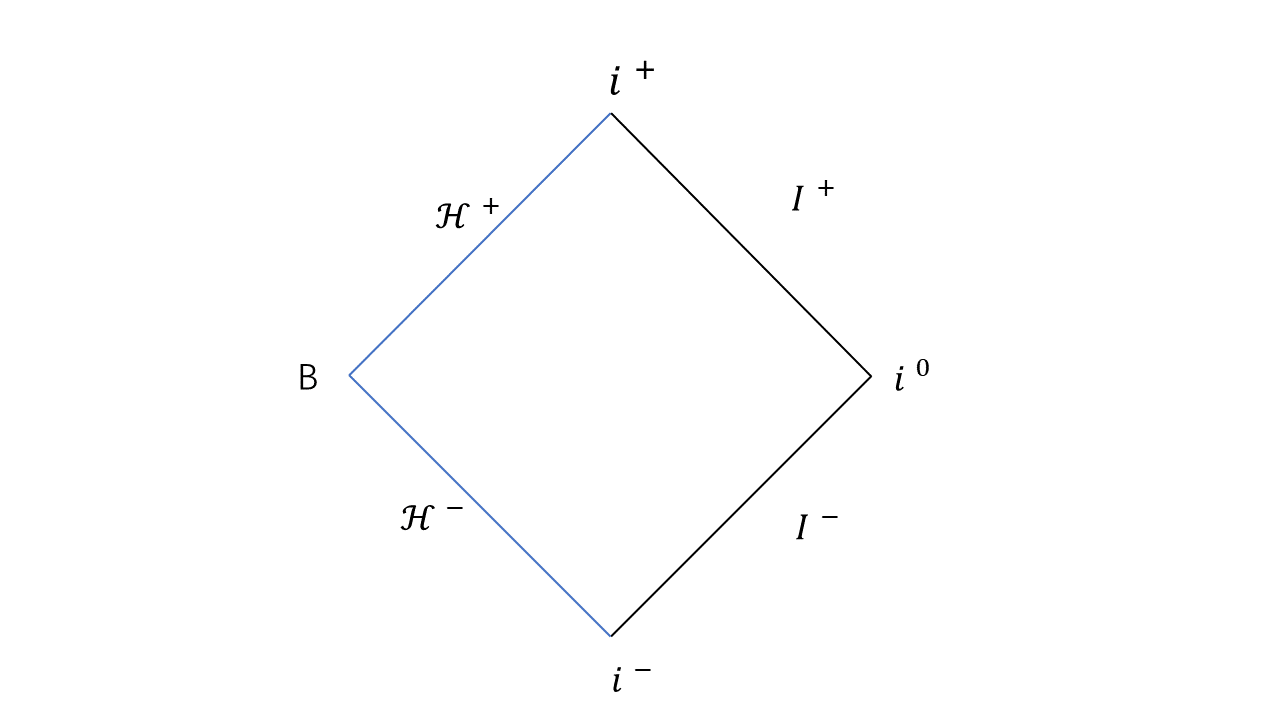}}
\caption{The Penrose diagram of the outside region of the Schwarzschild black hole.} \label{f1}
\end{figure}

We consider linearized gravitational theory coupled to a conserved stress tensor in the near horizon region of the Schwarzschild black hole. The line element of Schwarzschild black hole in the retarded coordinates $(u,r,x^A)$ is
\begin{align}
\label{metric}
    &\td s^2=-f(r)\td u^2-2\td u\td r+\Omega^2 \gamma_{AB}\td x^A \td x^B\,, \\
    &f(r)=\frac{r}{\Omega}\, , \quad \Omega=r+r_s\, .
\end{align}
Here $x^A=(z,\bz)$ are the complex stereographic coordinates which are related to the usual angular variables $(\theta,\phi)$ by $z=\cot\frac\theta2 e^{i\phi}$, and $r_s$ is the Schwarzschild radius which is related to the mass of the black hole by $r_s=2M$. The metric of unit sphere in this coordinate system is 
 \begin{align}
    \gamma_{AB}=\begin{pmatrix}
0 & \gamma _{z\overline{z}}\\
\gamma _{z\overline{z}} & 0
\end{pmatrix}\, , \ga=\frac{2}{(1+z\Bar{z})^2}\, .
\end{align}
The $r=0$ hypersurface is the horizon $\cH$ with topology $S^2\times\RR$. Actually, the retarded coordinates just cover half of the horizon $\cH^-$ as illustrated in Figure \ref{f1}. Hence we will only pay attention in this work to the $\cH^-$ part while everything can be similarly repeated on $\cH^+$ by introducing the advanced coordinates $(v,r,z,\bz)$. Those two parts are connected at the bifurcation point $B$. The near bifurcation symmetries derived in \cite{Adami:2020amw} provide a natural way of connecting the near horizon symmetries on $\cH^-$ and $\cH^+$.

Under a background metric $g_{\mu\nu}$, the action of a spin-2 theory of field $h_{\mu\nu}$ with zero cosmological constant is given by the quadratic part of the expansion of
the Einstein-Hilbert action 
\begin{align}
    S_{h}[h;g]&=\int \frac{\td^4x\sqrt{-g}}{4}\left[\frac{1}{2}(\nabla_\rho h)^2-\frac{1}{2}(\nabla_\rho h_{\mu\nu})^2+\nabla_\rho h_{\mu\nu}\nabla^\mu h^{\nu\rho}-\nabla_{\nu}h\nabla_{\mu}h^{\mu\nu}  \right]\, ,\label{action of spin-2} \\
    S_M[h;g]&=\int \frac{\td^4x\sqrt{-g}}{4} h^{\mu\nu}T_{\mu\nu}\, ,
\end{align}
where all indices are lowered and raised by $g_{\mu\nu}$ and $g^{\mu\nu}$ which is a solution of a vacuum Einstein field equation, $h=h_{\mu\nu}g^{\mu\nu}$, $\nabla_\mu$ is the covariant derivative corresponding to the background metric $g_{\mu\nu}$, and $T_{\mu\nu}$ is a generic conserved stress tensor satisfying $\nabla_\mu T^{\mu\nu}=0$, and coupled to the spin-2 gauge field. Inherited from the diffeomorphism invariance of Einstein-Hilbert action, the theory described by the action \eqref{action of spin-2} is invariant under the gauge transformations \cite{Barnich:2004ts}
\begin{align}
\delta_{\zeta} h_{\mu\nu}=\nabla_\mu\zeta_{\nu}+\nabla_\nu\zeta_{\mu}\, ,
\end{align}
with an arbitrary vector $\zeta^\mu(x)$.

The equations of motion of the linearized theory are the Pauli-Fierz equations in a covariant way,
\be\label{Einsteineqs}
E_{\mu\nu}=T_{\mu\nu}.
\ee
where
\be
E_{\mu\nu}\equiv\n_\mu\n_\nu h + \n^\tau\n_\tau h_{\mu\nu} - \n^\tau\n_\mu h_{\nu\tau} - \n^\tau \n_\nu h_{\mu\tau} - g_{\mu\nu}(\n^\tau\n_\tau h-\n^\tau\n^\rho h_{\tau\rho}),
\ee
and we are using natural units with $8\pi G_{\textrm{N}}=1\,$.

We will work in the adapted Newman-Unti gauge \cite{Newman:1962cia}, for which we set
\be
\begin{aligned}
&h_{rr}=h_{rz}=h_{r\bz}=h_{ru}=0\, ,\\
    &T_{rr}=T_{rz}=T_{r\bz}=T_{ru}=0.
\end{aligned}
\label{condition}
\ee
Here, we set all the radial components of the stress tensor $T_{r\mu}$ to zero to adapt to the gauge conditions of the perturbative metric \cite{Conde:2016rom}. This can be done using an auxiliary conserved symmetric 2-tensor as was detailed in Appendix \ref{stresstensor}. Since the linearized fields can be resided on the horizon in any form, there is no clear physical criteria to impose near horizon fall-off conditions. Here, we will not impose any fall-off condition for the linearized metric fields and derive the most general residual gauge transformations and the solution space. 

We decompose the perturbative metric and stress tensor in the NU gauge as follows
\begin{align}\label{decomposition}
h_{\mu\nu}=\begin{pmatrix}
\mathcal{C} & 0 & \mathcal{U}_{A}\\
0 & 0 & 0\\
\mathcal{U}_{A} & 0 & \mathcal{N }_{AB}+\gamma_{AB}\mathcal{T}/2
\end{pmatrix}\, , \quad T_{\mu\nu}=\begin{pmatrix}
T_{uu} & 0 & T_{uA}\\
0 & 0 & 0\\
T_{uA} & 0 & \mathbf{T}_{AB}+\gamma_{AB}\mathbf{T}/2
\end{pmatrix}\, , 
\end{align}
where $\mathcal{N }_{AB}$ satisfying $\mathcal{N }_{AB}\gamma^{AB}=0$ is the traceless part of $h_{AB}$, and $\mathcal{T}$ denotes the trace part. The same type of decomposition is also applied for $T_{AB}$. 

The gauge invariance of the linearized theory indicates that not all the equations of motion are independent. The constraints among them are inherited from the Bianchi identity of the full Einstein theory, which is now given by $\nabla_\mu (E^{\mu\nu}-T^{\mu\nu})=0$. Taking into account those constraints, we can arrange the ten equations of motion as follows:
\begin{itemize}
\item Four hypersurface equations:
\bea
E_{rr}=T_{rr},\quad E_{rz}=T_{rz},\quad E_{r\bz}=T_{r\bz},\quad E_{ru}=T_{ru},.
\eea
\item Two standard equations:
\bea
E_{zz}=T_{zz},\quad E_{\bz\bz}=T_{\bz\bz}.
\eea
\item One trivial equation:
\bea
E_{z\bz}=T_{z\bz}.
\eea
\item Three supplementary equations:
\bea
E_{uz}=T_{uz},\quad E_{u\bz}=T_{u\bz},\quad E_{uu}=T_{uu}.
\eea
\end{itemize}
\noindent
The advantage of such arrangement is well explained in the literature \cite{Bondi:1962px,Sachs:1962wk,Barnich:2010eb,Conde:2016rom}. Once the hypersurface equations and the standard equations are satisfied, the trivial equation is fulfilled automatically and all supplementary equations are left with only one order in the near horizon (small $r$) expansions.

\subsection{Near horizon symmetries}

The linearized gravity theory is invariant under the gauge transformation $\delta_\zeta h_{\mu\nu}=\nabla_\mu \zeta_\nu + \nabla_\nu \zeta_\mu$. We are interested in the residual gauge transformations that preserve the gauge conditions in \eqref{condition}. 

We start from the first constraint $h_{rr}=0$ which leads to $\partial_r\zeta_r=0$. Hence $\zeta_r=-\mathcal{F}$ where $\mathcal{F}$ is an arbitrary function of $(u,z,\bz)$. The minus sign is for later convenience. We continue with $h_{rA}=0$, which yields
\begin{align}
   \partial_r\zeta_A-\frac{2\zeta_A}{\Omega}=D_A\mathcal{F}\, .
\end{align}
One can easily read the general solution of $\zeta_A$ from the above equation,
\begin{align}
    \zeta_A=\Omega^2Y_A-\Omega D_A\mathcal{F}\, ,
\end{align}
where $Y_z, Y_{\bz}$ are arbitrary functions of $(u,z,\bz)$ and $D_A$ is the covariant derivative associated with the spherical metric $\gamma_{AB}$. Finally, from the last condition $h_{ru}=0$, we obtain
\begin{align}
    \partial_r\zeta_u=\partial_u\mathcal{F}-\frac{r_s\mathcal{F}}{\Omega^2}\, .
\end{align}
Hence, 
\begin{align}
    \zeta_u=r\partial_u\mathcal{F}-f\mathcal{F}-Z\, ,
\end{align}
where $Z$ is a arbitrary function of $(u,z,\bz)$.
The general solution of $\zeta$ in vector form 
by raising the index with $g^{\mu\nu}$ is:
\be
\zeta^\mu=\left[\mathcal{F},\, Z-r \p_u \mathcal{F}, \,Y^A-\frac{D^A\mathcal{F}}{\Omega}\right]. 
\ee
We have thus established the form of the residual gauge transformations of the linearized theory. Obviously, the near horizon symmetry includes both near horizon supertranslations and superrotations uncovered in \cite{Donnay:2015abr}.

The symmetry algebra is defined through $[\delta_{\zeta_1},\delta_{\zeta_2}]h_{\mu\nu}=\delta_{[\zeta_1,\zeta_2]_{_l}}h_{\mu\nu}$ . One can easily check that in the linearized case, the near horizon symmetry algebra is not the standard Lie algebra between vectors, but an Abelian algebra such that $[\zeta_1,\zeta_2]_{_l}=0$ where the subscript ``$l$'' denotes the algebra in linearized theory. This is another effect of linearisation.

The transformation law of the metric fields in the decomposition \eqref{decomposition} under the residual gauge transformation can be obtained easily as 
\begin{align}
    \delta_\zeta\mathcal{C}&=2r\partial_u^2\mathcal{F}+\left(r\partial_r f-2f\right)\partial_u\mathcal{F}-\partial_rfZ-2\partial_uZ\,, \label{variation C}\\
    \delta_\zeta\mathcal{U}_A&=-r_s\partial_uD_A\mathcal{F}-fD_A\mathcal{F}+\Omega^2\partial_u Y_A-D_AZ\, ,\label{variation U}\\
    \delta_\zeta \mathcal{T}&=-2\Omega D^AD_A\mathcal{F}-4r\Omega\partial_u\mathcal{F}+2\Omega^2D^AY_A+4\Omega Z\,, \label{variation trace}\\
    \delta_\zeta \mathcal{N }_{AB}&=2\Omega D_AY_B-2\Omega D_AD_B\mathcal{F}\, .\label{variation traceless}
\end{align}

\subsection{Solution space}

In this section, we want to solve the equations of motion in the near horizon region by assuming that the metric fields are given in a small $r$ expansion. We define a dimensionless quantity $\lambda=\frac{r}{r_s}$, and assume the metric and the stress tensor can be expanded in the form of series expansion as
\be
h_{\mu\nu}=h_{\mu\nu}^0+\sum_{n=1}^\infty\lambda^n h_{\mu\nu}^{(n)}\, , \quad T_{\mu\nu}=T_{\mu\nu}^0+\sum_{n=1}^\infty\lambda^n T_{\mu\nu}^{(n)}\, . 
\ee

We begin with the simplest hypersurface equation $E_{rr}=0$, which yields
\begin{align}\label{hyper rr}
0=\partial_\lambda^2\left(\frac{\mathcal{T}}{\xi}\right)\, ,
\end{align}
where we introduce $\xi=(1+\lambda)$ for notational convenience. The solution for the trace part is
\begin{align}
\mathcal{T}=\mathcal{T}^0+\lambda\mathcal{T}^{(1)}+\lambda^2(\mathcal{T}^{(1)}-\mathcal{T}^{0})\, ,
\end{align}
which implies that, under the NU gauge, the trace part of spin-2 field truncates in the small $r$ expansion.

Then, the second group $E_{rA}=0$ leads to
\begin{align}\label{hyper-rA}
    \frac{1}{r_s^2\xi^2}\p_\lambda\left[\xi^4\p_\lambda\left(\frac{\mathcal{U}_A}{\xi^2}\right)\right]-\frac{1}{ r_s^3}\p_\lambda\left(\frac{D^B\mathcal{N }_{BA}}{\xi^2}\right)-\frac{1}{ r_s^3}\p_\lambda\left(\frac{D_A\mathcal{T}}{2\xi^2}\right)=0\, .
\end{align}
Considering $\mathcal{N }_{AB}$ and $\mathcal{T}$ as known functions, $\mathcal{U}_A$ is completely fixed up to four integration constants of $r$ which are the two leading orders $\mathcal{U}_A^{(0)}$ and $\mathcal{U}_A^{(1)}$. Suppose that $\mathcal{N }_{AB}$ is given as initial data as
\be
\mathcal{N }_{AB}=\sum_{m=0}^\infty\mathcal{N }_{AB}^{(m)} r^m.
\ee
The higher orders in the expansion of $\mathcal{U}_A$ for $n\ge 0$ are determined as follows
\begin{equation}
    \begin{aligned}
    &\mathcal{U}_{A}^{(n+2)}=\frac{1}{(n+2)(n+1)} \left\{2\sum_{m=0}^n(-1)^m (1+m)\mathcal{U}_{A}^{(n-m)}
   \right.\\
    &\left.+\sum_{m=0}^n\frac{(-1)^{m}(1+m)}{r_s}\left[(n+1-m)D^B\mathcal{N }_{BA}^{(n+1-m)}-(2+m)D^B\mathcal{N }_{BA}^{(n-m)}\right]\right. \\
    &\left.+\sum_{m=0}^n\frac{(-1)^{m}(1+m)}{2r_s}\left[(n+1-m)D_A\mathcal{T}^{(n+1-m)}-(2+m)D_A\mathcal{T}^{(n-m)}\right]\right\},
\end{aligned}
\end{equation}
where the series expansion on the right hand side of the above relation is from that the expansion of $\frac{1}{\xi}$ in powers of $\lambda$ does not truncate, which is 
\begin{align}
     \frac{1}{\xi^n}=\sum_{m=0}^{\infty}(-1)^m\frac{\Gamma(m+n)}{\Gamma(m)\Gamma(1+n)}\lambda^m\, .
\end{align}
This is the main difference of the solution space compared to the null infinity case \cite{Conde:2016rom}. An equivalent fact is that linearisation from the full theory is more complicated on a curved background than a flat spacetime background. Consequently, there is no advantage to obtaining the solution space of the linearized theory in Schwarzschild spacetime from the full theory in \cite{Adami:2021nnf}. This is the main reason we study the linearized theory from scratch in this work.\footnote{See also \cite{Aggarwal:2023qwl} for the study of the linearized theory in Schwarzschild spacetime in radiation gauge.}

We continue with the last hypersurface equation $E_{ru}=0$, which gives
\begin{equation}
    \begin{aligned}
   -\frac{2\partial_\lambda\left(\xi \mathcal{C}\right)}{r_s^2\xi^2}&+\frac{\partial_\lambda\left(\xi^2 D^A\mathcal{U}_A\right)}{r_s^3\xi^4}-\frac{D^AD^B\mathcal{N }_{AB}}{r_s^4\xi^4}-\frac{D^AD_A\mathcal{T}}{2r_s^4\xi^4}\\
   &\quad\quad-\frac{\partial_u\p_\lambda \mathcal{T}}{r_s^3\xi^2 }+\frac{f\mathcal{T}}{r_s^4\xi^4}+\frac{f}{r_s^4\xi}\partial_\lambda\left(\frac{\partial_\lambda \mathcal{T}}{\xi}\right)+\frac{\partial_\lambda \mathcal{T}}{2r_s^4\xi^4}=0\, ,
\end{aligned}
\end{equation}
which controls $\mathcal{C}$ up to an integration constant at the leading order $\mathcal{C}^{(0)}$. The higher orders are determined by
\begin{align}
\mathcal{C}^{(1)}&=\frac{2D^A\mathcal{U}_{A}^0+D^A\mathcal{U}_{A}^{(1)}}{2r_s}-\frac{D^AD^B\mathcal{N }_{AB}^0}{2r_s^2}-\frac{D^AD_A\mathcal{T}^0}{4r_s^2}-\frac{\partial_u\mathcal{T}^{(1)}}{2r_s}+\frac{\mathcal{T}^{(1)}}{4r_s^2}-\mathcal{C}^0\, ,
\end{align}
and
\begin{equation}
    \begin{aligned}
    &\mathcal{C}^{(n+1)}=\frac{1}{n+1}\left\{\sum_{m=0}^n(-1)^{m+1}\mathcal{C}^{(n-m)}\right.\\
    &+\sum_{m=0}^n\frac{(-1)^{m+1}(1+m)(2+m)}{4r_s^2}\left[D^AD^B\mathcal{N }_{AB}^{(n-m)}+D^AD_A\mathcal{T}^{(n-m)}/2\right]\\
    &+\sum_{m=0}^n\frac{(-1)^m}{2r_s}\left[(n+1-m)D^A\mathcal{U}_A^{(n+1-m)}+2(1+m)D^A\mathcal{U}_A^{(n-m)}\right]\\
    & -\sum_{m=0}^n\frac{(-1)^m(n+1-m)}{2r_s}\partial_u\mathcal{T}^{(n+1-m)}\\
    &+\sum_{m=0}^{n}\frac{(-1)^m(1+m)(2+m)(n+1-m)}{8r_s^2}\mathcal{T}^{(n+1-m)}\\
    &+\sum_{m=0}^{n-1}\frac{(-1)^m(1+m)(2+m)}{12r_s^2}\\
    &\quad  \times
        \left[\frac{6(n+1-m)(n-m)}{(2+m)}\mathcal{T}^{(n+1-m)}-3(n-m)\mathcal{T}^{(n-m)}+(3+m)\mathcal{T}^{(n-1-m)}\right]\bigg\}\, ,
\end{aligned}
\end{equation}
 for $n\ge 1$.

We now turn to the standard equations. The standard equation $E_{zz}=T_{zz}$ is reduced to
\begin{align}\label{standard}
    \partial_u \partial_\lambda\mathcal{N }_{AB}=\partial_\lambda D_A \mathcal{U}_B+\frac{\partial_\lambda \mathcal{N }_{AB}}{2r_s\xi^2}+\frac{\lambda \xi}{2r_s}\partial^2_\lambda\left(\frac{\mathcal{N }_{AB}}{\xi}\right)-\frac{\mathcal{N }_{AB}}{r_s\xi^3}+\frac{\partial_u\mathcal{N }_{AB}}{\xi}-\frac{r_s}{2}\mathbf{T}_{AB}\, .
\end{align}
This equation controls the time evolution of the initial data $\mathcal{N }_{AB}$ except its leading order $\mathcal{N }_{AB}^{(0)}$. Following the terminology of \cite{Bondi:1962px,Adami:2021nnf}, we refer to the time derivative of the components at this order as \textit{news} functions. The precise information that is extracted from~\eqref{standard} is
\begin{align}
    \partial_u \mathcal{N }_{AB}^{(1)}=D_A\mathcal{U}_{B}^{(1)}+\partial_u \mathcal{N }_{AB}^0+\frac{\mathcal{N }_{AB}^{(1)}-2\mathcal{N }_{AB}^0}{2r_s}-\frac{\mathbf{T}^{0}_{AB}}{2r_s}\, , 
\end{align}
and, for $n\ge 1$,
\begin{equation}
    \begin{aligned}
        \partial_u &\mathcal{N }_{AB}^{(n+1)}=D_A\mathcal{U}_{B}^{(n+1)}+\frac{1}{(n+1)}\left\{\sum_{m=0}^n(-1)^m\partial_u \mathcal{N }_{AB}^{(n-m)}-\frac{\mathbf{T}_{AB}^{(n)}}{2r_s}\right.\\
        &+\sum_{m=0}^n\frac{(-1)^m(1+m)}{2r_s}\left[(n+1-m)\mathcal{N }_{AB}^{(n+1-m)}-(2+m)\mathcal{N }_{AB}^{(n-m)}\right]
        \\
        &+\sum_{m=0}^{n-1}\frac{(-1)^m}{r_s}\left[(1+m)\mathcal{N }_{AB}^{(n-1-m)}-(n-m)\mathcal{N }_{AB}^{(n-m)}\right]\\
        &\left.+\sum_{m=0}^{n-1}\frac{(n-m)(n+1-m)}{2r_s}\mathcal{N }_{AB}^{(n+1-m)}\right\}\, .
    \end{aligned}
\end{equation}

When the hypersurface equations and standard equations are satisfied, the trivial equation is satisfied automatically, which can be verified from the $r$ component of the Bianchi identity $\nabla^\mu(E_{\mu\nu}-T_{\mu\nu})=0$, from which one can obtain
\be
\frac{2(E_{z\bz}-T_{z\bz})}{\Omega^2\ga}=0\, .
\ee

For the supplementary equations, only one order in the expansion is involved. Three supplementary equations fix the time evolution of the integration constants as
\begin{align}
    &\frac{2\partial_u\mathcal{C}^0}{r_s}+\frac{D^AD_A\mathcal{C}^0}{r_s^2}-\frac{2\partial_u\left(D^A\mathcal{U}_A^0\right)}{r_s^2}-\frac{D^A\mathcal{U}_A^0}{r_s^3}+\frac{\left(2\partial_u^2 \mathcal{T}^0+\partial_u \mathcal{T}^0\right)}{2r_s^2}=T_{uu}^0\, , \label{Cu}\\
    &\frac{2r_s\partial_u \mathcal{U}_{A}^0-r_s\partial_u\mathcal{U}_{A}^{(1)}+2\mathcal{U}_{A}^0}{r_s^2}+\frac{D_A\mathcal{C}^{(1)}}{r_s}-\frac{\partial_uD^B\mathcal{N }_{AB}^0}{r_s^2}\notag\\
    &\hspace{6cm}+\frac{D^BD_B\mathcal{U}_{A}^0-D^BD_A\mathcal{U}_{B}^0}{r_s^2}=T_{uA}^0\, .\label{Nu}
\end{align}
Substituting $\mathcal{C}^{(1)}$ in the second equation yields
\begin{equation}
    \begin{aligned}
    &\frac{2r_s\partial_u \mathcal{U}_{A}^0-r_s\partial_u\mathcal{U}_{A}^{(1)}+\mathcal{U}_{A}^0}{r_s^2}+\frac{D_A\mathcal{T}^{(1)}}{4r_s^3}-\frac{\partial_uD_A\mathcal{T}^{(1)}}{2r_s^2}-\frac{D_AD^BD_B\mathcal{T}^0}{4r_s^3}\\    &+\frac{2D^BD_B\mathcal{U}_{A}^0+D_AD^B\mathcal{U}_B^{(1)}}{2r_s^2}-\frac{D_A\mathcal{C}^0}{r_s}-\frac{D_AD^BD^C\mathcal{N }_{BC}^0}{2r_s^3}-\frac{\partial_uD^B\mathcal{N }_{AB}^0}{r_s^2}=T_{uA}^0\, .
\end{aligned}
\end{equation}

\subsection{Near horizon surface charges}

For the (linearized) Einstein gravity, the conserved charge defined on bifurcation 2-sphere $B$ is given by \cite{Barnich:2001jy}
\begin{equation}
    {\mathcal{Q}}_\zeta =\int_B\,\sqrt{-g} \Big( h^{\lambda [ u} \nabla _{\lambda} \zeta^{r]} - \zeta^{\lambda} \nabla^{[u} {h^{r]}}_{\lambda} - \frac{1}{2} h \nabla ^{[u} \zeta^{r]} + \zeta^{[u} \nabla _{\lambda} h^{r] \lambda} - \zeta^{[u} \nabla^{r]}h \Big)\td z \td \bz.
\end{equation}
Inserting the solution space and the near horizon symmetries, one obtains 
\begin{multline}\label{chargefull}
{\mathcal{Q}}_{\zeta(\mathcal{F},Z,Y^A)}=\int_B \sqrt{-g}\left\{\frac{\mathcal{F}\mathcal{C}^0}{r_s}+\frac{\mathcal{U}_A^{(1)}D^A\mathcal{F}}{2r_s^2}+\frac{Y^A\left(2\mathcal{U}_A^{0}-\mathcal{U}_A^{(1)}\right)}{2r_s}\right.\\
\left.\frac{Z\left(\mathcal{T}^{0}-\mathcal{T}^{(1)}\right)}{2r_s^3}+\frac{\mathcal{F}\mathcal{T}^{0}}{2r_s^3}+\frac{\mathcal{F}\partial_u\mathcal{T}^{0}-\partial_u\mathcal{F}\mathcal{T}^{0}}{r_s^2} \right\}\mathrm{d}z\mathrm{d}\bz\, .
\end{multline}
The charge algebra of the linearized theory is defined by
\begin{align}
[\mathcal{Q}_{\zeta_1},\mathcal{Q}_{\zeta_2}]\equiv\delta_{\zeta_1}\mathcal{Q}_{\zeta_2}=\mathcal{Q}_{[\zeta_1,\zeta_2]_l}+\mathcal{K}_{\zeta_1,\zeta_2}\, ,
\end{align}
where $\mathcal{K}_{\zeta_1,\zeta_2}$ denotes a possible 2-cocycle term. To verify the charge algebra, one needs to apply the transformation law of the metric fields at the leading order,
\begin{align}
    \delta_{\zeta}\mathcal{C}=&-Z/r_s-2\partial_uZ+\mathcal{O}(\lambda)\,,\label{variation expansion C}\\
    \delta_\zeta\mathcal{U}_A=&-D_AZ+r_s^2\partial_uY_A-r_s\partial_uD_A\mathcal{F}+\lambda(-D_A\mathcal{F}+2r_s^2\partial_uY_A)+\mathcal{O}(\lambda^2)\, ,\label{variation expansion U}\\
    \delta_\zeta \mathcal{T}=&2r_s^2D^AY_A-2r_sD^AD_A\mathcal{F}+4r_sZ\notag\\
    &\quad\quad\quad+\lambda(4r_s^2D^AY_A-2r_sD^AD_A\mathcal{F}-4r_s^2\partial_u\mathcal{F}+4r_sZ)+\mathcal{O}(\lambda^2)\, ,\\
    \delta_\zeta \mathcal{N }_{AB}=&2r_s^2D_AY_B-2r_sD_AD_B\mathcal{F}+\mathcal{O}(\lambda)\, .
\end{align}
Inserting field variation generated by $\zeta^\mu_2(\mathcal{F}_2,Z_2,Y^A_2)$ for the linearized theory, the algebra of the charge is given by
\begin{equation}
\begin{aligned}
\delta _{\zeta _{2}}\mathcal{Q}_{\zeta _{1}} =&\mathcal{K}_{\zeta_1,\zeta_2}\\
=&\int_B\sqrt{-g}\left\{\frac{Z_{1} \partial _{u}\mathcal{F}_{2} +\mathcal{F}_{1} \partial _{u} Z_{2}}{r_{s}/2} +\frac{D^{A}\mathcal{F}_{1} \partial _{u} D_{A}\mathcal{F}_{2}}{r_{s}} \right. \\
 & \quad\quad\quad\left.+D_{A} Y_{1}^{A} \partial _{u}\mathcal{F}_{2} -\frac{Z_{1} D_{A} Y_{2}^{A}}{r_{s}} +\frac{Y_{1}^{A} D_{A}\mathcal{F}_{2}}{2r_{s}}-( 1\leftrightarrow 2)\right\}\mathrm{d}z\mathrm{d}\bz\, .
\end{aligned}
\end{equation}

\section{Near horizon soft graviton theorem }
\label{soft}

In this section we will derive a soft graviton theorem from the Ward identity of the near horizon symmetry. The success of deriving soft theorem from asymptotic/near horizon symmetry resides in a natural split of the symmetry charges into soft and hard parts \cite{Strominger:2013lka,Strominger:2013jfa,He:2014laa,He:2014cra,He:2015zea,Strominger:2017zoo,Cheng:2022xyr,Cheng:2022xgm,Mao:2023rca}. Unfortunately, it is completely unclear how one can organize the near horizon charge \eqref{chargefull} derived in previous section into soft and hard parts. Nevertheless, we can impose proper near horizon fall-off conditions to reduce the solution space and near horizon charges, such that a natural split will emerge.

After solving the hypersurface and standard equations, the solution space is spanned by a collection of boundary initial data. This set includes $\{\mathcal{N }_{AB}^{0},\mathcal{C}^0,\mathcal{U}_A^{0},\mathcal{U}_A^{(1)},\mathcal{T}^{0},\mathcal{T}^{(1)}\}$. Among these, the two dimensional trace-less tensor $\mathcal{N}_{AB}^{0}$ represents the propagating degree of freedom associated with a graviton in the bulk which are the flux falling into the horizon. The scalar field $\mathcal{C}^{0}$ represents the energy of the linearized theory, see also \cite{Adami:2021kvx,Adami:2023fbm} from a thermodynamical perspective. A two dimensional vector from the combination of $\mathcal{U}_{A}^{0}$and $\mathcal{U}_A^{(1)}$ represents the angular momentum perspective. The time evolution of those three quantities are controlled by the supplementary equations. There are still four free functions or degrees of freedom which consist of $\mathcal{T}^{0}, \mathcal{T}^{(1)}$ and one two dimensional vector from another type of combination of $\mathcal{U}_{A}^{0}$and $\mathcal{U}_A^{(1)}$. They are the fake propagating degree of freedom \cite{Adami:2021nnf} which can be eliminated by either imposing fall-off conditions or  redefining the symmetry parameters by involving field dependence \cite{Adami:2020ugu,Ruzziconi:2020wrb,Adami:2021sko}. We will use the fake degree of freedom to reorganize the solution space and recover a natural split of the charge.

\subsection{Fall-off conditions and charge}

According to \eqref{hyper rr}, we can completely turn off all orders of the trace component $\mathcal{T}$ by imposing the fall-off condition $\mathcal{T}=\mathcal{O}(\lambda^2)$. Due to the absence of $\mathcal{T}^0$ and $\mathcal{T}^{(1)}$, both the conserved charge and the equations of motion undergo a significant simplification. To incorporate with this fall-off requirement, the symmetry parameters are constrained by $\delta_\zeta \mathcal{T}=0$. From Eq. \eqref{variation trace}, we obtain
\begin{align}
\left(-D^AD_A\mathcal{F}+r_sD^AY_A+2Z\right)+\lambda\left(-2r_s\partial_u\mathcal{F}+r_sD^AY_A\right)=0\, .
\end{align}
Hence,
\begin{align}
    Z&=\frac{1}{2}D^AD_A\mathcal{F}-\frac{r_s}{2}D^AY_A\, ,\label{Z in YF}\\
    \mathcal{F}&=T+\frac{1}{2} D^A\mathbf{Y}_A\, , \label{F in Y}
\end{align}
where $T$ is an arbitrary function of $(z,\bz)$ and 
\begin{align}
    \mathbf{Y}_A=\int\td u Y_A\, .
\end{align}Here $T(z,\bz)$ generates the supertranslation of linearized gravitational fields on the horizon, and $Y_A$ is the generalized near horizon superrotation.
We are mainly interested in the soft theorem associated to the supertranslation charge, which in the reduced solution space becomes
\begin{align}\label{charge T 1}
    Q_T&=\int_B \ga T\left(r_s\mathcal{C}^0-\frac{D^A\mathcal{U}_A^{(1)}}{2}\right)\td^2 z\, .
\end{align}
The supertranslation charge can be evaluated on the horizon applying the relations in \eqref{Cu} and \eqref{Nu},
\begin{equation}
    \begin{aligned}
        Q_T=&\frac{1}{2r_s}\int_{\mathcal{H}^-} \ga T\partial_u\left(D^AD^B\mathcal{N}_{AB}^0\right)\td^2 z\td u\\
        &+\frac{r_s}{2}\int_{\mathcal{H}^-} \ga T\left(r_sT_{uu}^0+D^AT_{uA}^{0}\right)\td^2 z\td u\\
        &-\frac{1}{2}\int_{\mathcal{H}^-}\ga T\left(D^A\mathcal{U}_A^0/r_s+D^AD_A\mathcal{C}^0+D^AD_A\mathcal{C}^{(1)}\right)\td^2 z\td u\, .
\end{aligned}
\end{equation} 
The first line of this charge expression has exactly the desired forms as a soft part of the charge, and the second line is in the form of a hard part. The third line can be turned off by imposing proper fall-off conditions, for which we set
\begin{align}
   \mathcal{U}_A^0+r_sD_A\left(\mathcal{C}^0+\mathcal{C}^{(1)}\right)=0\, .
\end{align}
In order to take the symmetry parameter $\zeta^\mu(\mathcal{F},Y^A)$ to be compatible with this fall-off condition, those parameters should satisfy 
\begin{align}
    \delta_\zeta\left[ \mathcal{U}_A^0+r_sD_A\left(\mathcal{C}^0+\mathcal{C}^{(1)}\right)\right]=0.\label{condition removing non-soft/hard parts}
\end{align}
From \eqref{variation C} and \eqref{variation U}, we have 
\begin{align}
    \delta_{\zeta}\mathcal{U}_A^0&=-D_AZ+r_s^2\partial_uY_A-r_s\partial_uD_A\mathcal{F}\, ,\\
    \delta_\zeta \left[r_sD_A\left(\mathcal{C}^0+\mathcal{C}^{(1)}\right)\right]&=D_AZ-2r_s\partial_uD_AZ+2r_s^2\partial^2_uD_A\mathcal{F}-r_s\partial_uD_A\mathcal{F}\, .
\end{align}
Then, Eq. \eqref{condition removing non-soft/hard parts} could be reorganized as 
 \begin{align}\label{constriant charge}
     \partial_u\left[r_s\left(\gamma_{AB}+2D_AD_B\right)Y^B-(2\gamma_{AB}+D_AD_B)D^B\mathcal{F}\right]=0\,.
 \end{align}
Clearly, a pure supertranslation with $\mathcal{F}=T$ and $Y^A=0$ satisfies this constraint. Finally, the near horizon supertranslation charge can be split as
\begin{align}
    Q_T=Q^S+Q^H\, ,
\end{align}
where $Q^{S},Q^{H}$ denote the soft and hard charges respectively,
\begin{align}
   Q^S&=\frac{1}{2r_s}\int_{\mathcal{H}^-} \ga T\partial_u\left(D^AD^B\mathcal{N}_{AB}^0\right)\td^2 z\td u\, , \\
   Q^H&=\frac{r_s}{2}\int_{\mathcal{H}^-} \ga T\left(r_sT_{uu}^0+D^A T_{uA}^{0}\right)\td^2 z\td u\, .\label{hard}
\end{align}
In the null infinity analysis \cite{Strominger:2013jfa,He:2014laa}, it is normally assumed that there is no long-range magnetic mass aspect. This assumption can be adapted to the near horizon case by imposing
\begin{align}
    \left[D_zD_z\mathcal{N}^0_{\bz\bz}-D_{\bz} D_{\bz}\mathcal{N}^0_{zz}\right]_{\mathcal{H}^{-}_{\pm}}=0\, ,
\end{align}
Then we can rewrite the soft charge as 
\begin{align}\label{soft supertranslation charge}
    Q^{S}=-\frac{1}{r_s}\int_{\mathcal{H}^-}  \partial_{\bz}T\partial_u\left(\frac{\partial_{\bz}\mathcal{N}_{zz}^0}{\ga}\right)\td^2 z\td u\, ,
\end{align}
For the hard part of the charge, it consists only the contribution from the stress tensor. We have used an auxiliary field to modify the stress tensor to make it more consistent with our gauge conditions. Basically, it will just change the way of coupling the matter fields to gravity. In the null infinity case, the modification of the stress tensor does not affect the leading soft theorem derived from the Ward identity of the supertranslation \cite{Conde:2016rom} which simply reflects the fact that the leading soft graviton theorem is universal for any type of gravitational theory \cite{Elvang:2016qvq}. In contrast, the hard part of the near horizon supertranslation charge is sensitive to the modification of the stress tensor. Correspondingly, this indicates that the soft theorem derived from the Ward identity of the near horizon supertranslation charge will be dependent on the way of the matter fields couplings. In this work, we are interested in the case where the near horizon supertranslation charge has a similar expression as the null infinity case. We will further use the freedom in the auxiliary fields to turn off $T_{uA}^{0}$, and the precise modification is presented in Appendix \ref{stresstensor}. Thus, the hard part of the charge becomes
\be
Q^H=\frac{r_s}{2}\int_{\mathcal{H}^-} \ga T\left(r_sT_{uu}^0\right)\td^2 z\td u\, .
\ee

In Section \ref{soft theorem coordinates}, we will demonstrate that the soft charge creates a low-energy soft graviton in near horizon states, and the action of the hard charge in those states leads to the soft factor, which is akin to the scenario in flat spacetime.

\subsection{Commutation relations}

Now we derive the commutation relation between the supertranslation charge and gauge field using the covariant phase space approach. Following the standard procedure in \cite{Lee:1990nz,Iyer:1994ys}, one can find the presymplectic potential of a free spin-2 theory defined in \eqref{action of spin-2} as follows
\begin{equation}
    \begin{aligned}
    \boldsymbol{\Theta}=\frac{1}{2}\bigg(\frac{1}{2}\delta h\nabla^\rho h&-\frac{1}{2}\delta h\nabla^\mu h_{\mu}^{\ \rho}-\frac{1}{2}\delta h_{\mu}^{\ \rho}\nabla^\nu h \\
    &\left.-\frac{1}{2}\delta h^{\mu\nu}\nabla^\rho h_{\mu\nu}+\delta h^{\mu\nu}\nabla_\mu h_{\nu}^{\ \rho}\right)\sqrt{-g}\td x_{\rho}\,,
\end{aligned}
\end{equation}
where $\td x_\rho=\frac{1}{3!}\hat{\epsilon}_{\rho\mu\nu\sigma}\td x^\mu\wedge\td x^\nu\wedge\td x^\sigma$. The presymplectic form $\boldsymbol{\omega }$ is defined by
\begin{align}
    \boldsymbol{\omega }\equiv \delta \boldsymbol{\Theta}\, .
\end{align}
Note that, in our case, the Cauchy surface is $\mathcal{H}^{-}$, and thus after imposing near horizon fall-off conditions and gauge fixing, the presymplectic form is reduced to
\begin{align}
\boldsymbol{\omega }_{uz\bz}|_{\mathcal{H}^-}&=\frac{\ga \gamma^{AC}\gamma^{BD}}{4r_s^2}\delta\partial_u\mathcal{N}^{0}_{AB}\wedge \delta\mathcal{N}^0_{CD}\, .
\end{align}
We can write down the canonical commutator from the presymplectic form as
\begin{align}
    [\partial_u\mathcal{N}^0_{AB}(u,z,\bz),\mathcal{N}_{CD}^0(g,w,\bar{w})]=-i\frac{4r_s^2\gamma_{AC}\gamma_{BD} }{\ga}\delta(u-g)\delta^2(z-w)\, .
\end{align}
After a direct computation, we arrive at
\begin{align}
    [Q_T,\mathcal{N}_{AB}^0]=[Q^S,\mathcal{N}_{AB}^0]=i\left(-2r_sD_AD_BT\right)\, .
\end{align}
As expected, the soft charge generates the supertranslation. A similar computation can be done when matter fields coupled. For instance, for a massless scalar field as described in Appendix \ref{stresstensor}, one has 
\begin{align}
    [\partial_u\bar{\phi}(u,z,\bz),\phi(g,z,\bz)]&=-\frac{i}{r_s^2\ga}\delta(u-g)\delta^2(z-w)\, ,\\
    [Q_T,\phi]=[Q^H,\phi]&=-\frac{i}{2}T\partial_u\phi\, .
\end{align}
The second line above in the momentum space can be written as 
\begin{align}\label{commutation hard}
    [Q_T,\phi_k]=\frac{E_k}{2}T\phi_k\, .
\end{align}

\subsection{Mode expansion}

We will now derive the mode expansions of the near horizon gauge fields, which are the key step for formulating a soft theorem from near horizon symmetry. It is convenient to write down the mode expansion of the free field operators in the isotropic coordinates $(t,x_1,x_2,x_3)$, in which the line element of the Schwarzschild spacetime is
\begin{align}
    \td ^2s=-\left(\frac{4\rho-r_s}{4\rho+r_s}\right)^2\td^2 t+\left(\frac{4\rho+r_s}{4\rho}\right)^4\td \Vec{x}\cdot\td\Vec{x}\, .
\end{align}
The coordinate transformations from $(u,r,z,\bz)$ to $(t,x_1,x_2,x_3)$ are 
\begin{align}
    &t=u+\Omega+r_s\ln\left(\frac{r}{r_s}\right)\, ,\\
    &x_1=\rho\frac{z+\bz}{z\bz+1}\, , \ x_2=\rho\frac{z-\bz}{i(z\bz+1)}\, , \,  x_3=\rho\frac{z\bz-1}{z\bz+1}\, ,
\end{align}
where $\rho$ is a function of $r$, defined by
\begin{align}
    \rho=\frac{1}{4}\left[2r+r_s+2\sqrt{r(r+r_s)}\right]\, ,
\end{align}
thus the near horizon limit is $\rho\rightarrow \frac{r_s}{4}$. 

The external Schwarzschild spacetime is static, with $\partial_t$ serving as a timelike Killing vector everywhere which leads to a definition of positive energy of a moving particle as $\omega=-p_0$. In the isotropic coordinates, the dispersion relation for massless particles is 
\begin{align}
    -\left(\frac{4\rho+r_s}{4\rho-r_s}\right)^2 \omega ^{2} +\left(\frac{4\rho}{4\rho+r_s}\right)^4\vec{p}^{2}=0\, .
\end{align}
The covariant measure of one-particle phase space for the mode expansion is
\begin{align}
\int\frac{\mathrm{d}\omega\mathrm{d}^3\vec{p}}{(2\pi)^3}\delta(p^\mu p_\mu)\Theta(\omega)=\left(\frac{4\rho-r_s}{4\rho+r_s}\right)^2\int\frac{\mathrm{d}^3\vec{p}}{(2\pi)^3 2\omega}\, .
\end{align}
The mode expansion of a free graviton field is given by
\begin{align}
    h_{\mu\nu}=\sum_{\alpha=\pm}\left(\frac{4\rho-r_s}{4\rho+r_s}\right)^2\int\frac{\td^3 p}{(2\pi)^3}\frac{1}{2\omega}\left[\epsilon^{\alpha\star}_{\mu\nu }a_{\alpha}(\vec{p})e^{i\Vec{p}\cdot\Vec{x}}+\epsilon^{\alpha}_{\mu\nu }a^{\dagger}_{\alpha}(\vec{p})e^{-i\Vec{p}\cdot\Vec{x}}\right]\, ,
\end{align}
where $\epsilon_{\mu\nu}^\alpha$ is a product of the pair polarization vectors  $\epsilon_{\mu\nu}^\alpha=\epsilon^\alpha_\mu\epsilon^\alpha_\nu$.
We will use a new set of null parametrizations for describing a massless particle. The momentum of a graviton can be parameterized as follows:
\begin{equation}
\begin{aligned}\label{graviton momentum}
p_\mu=\frac{\omega}{1+z\bz}&\frac{(4\rho+r_s)^3}{16\rho^2(4\rho-r_s)}\\
&\times\left[\frac{16\rho^2(r_s-4\rho)}{(4\rho+r_s)^3}(1+z\bz),(z+\bz),-i(z-\bz),(1-z\bz)\right]\, .
\end{aligned}
\end{equation}
Correspondingly, the polarization vectors that are perpendicular to the direction of
propagation of graviton are reparametrized as 
\begin{align}
    \epsilon^{+\mu}&=\frac{1}{\sqrt{2}}\left(\frac{4\rho}{4\rho+r_s}\right)^2\left[\frac{(4\rho+r_s)^3}{16\rho^2(4\rho-r_s)}\bz,1,-i,z\right]\, ,\\
    \epsilon^{-\mu}&=\frac{1}{\sqrt{2}}\left(\frac{4\rho}{4\rho+r_s}\right)^2\left[\frac{(4\rho+r_s)^3}{16\rho^2(4\rho-r_s)}z,1,i,\bz\right]\, ,
\end{align}
with the following properties
\begin{align}
    \epsilon^\alpha_{r}=0\, ,\, p^\mu\epsilon^{\alpha}_{\mu}=0\, ,\, \epsilon^{\alpha\mu}\epsilon^{\star\beta}_{\mu}=\delta^{\alpha\beta}\, .
\end{align}
Projecting those two vectors on the sphere by $\epsilon^{\pm }_A=\frac{\partial x^\mu}{\partial x^A}\epsilon^{\pm}_\mu$, we obtain
\begin{align}
    \epsilon_A^+=\delta_{zA}\frac{(4\rho+r_s)^2}{4\sqrt{2}\rho(1+z\bz)}\, , \, \epsilon_{A}^{-}=\delta_{\bz A}\frac{(4\rho+r_s)^2}{4\sqrt{2}\rho(1+z\bz)}\, .
\end{align}
Here, we introduce the parameter $R = \rho - \frac{r_s}{4}$ to position the horizon at $R\rightarrow 0$. Following the treatment used in \cite{Cheng:2022xyr,Cheng:2022xgm}, we implement a near-horizon regularization by replacing $R$ with $R + i\mathcal{R}$ to regularize the divergence from the near-horizon limit. Eventually, the near
horizon field is related to the plane wave modes by
\begin{equation}\label{mode expansion}
    \begin{aligned}
    \mathcal{N}^0_{zz}=&\ga\frac{8}{\pi^2}\frac{(r_s+2R)^6}{(r_s+4R)^5}\int\td\omega\left(\frac{4R^2}{r_s^2+2r_sR}\right)^{-ir_s\omega}\\
    &\times \sin\left[\frac{2\omega(R+r_s/2)^3}{R(2R+r_s/2)}\right]\left(a_+e^{-i\omega u-i\omega\frac{2(R+r_s/2)^2}{2R+r_s/2}}+a_-^\dagger e^{i\omega u+i\omega\frac{2(R+r_s/2)^2}{2R+r_s/2}}\right)\, .
\end{aligned}
\end{equation}

\subsection{Soft graviton theorem in coordinates space}\label{soft theorem coordinates}

In this subsection, we derive the soft graviton theorem by applying the Ward identity of supertranslation
\begin{align}\label{ward identity}
    \left\langle {\rm out}\right|[Q_T,\mathcal{S}]\left|{\rm in}\right\rangle=0\Rightarrow Q^S\left|{\rm in}\right\rangle=-Q^H\left|{\rm in}\right\rangle\, ,
\end{align}
where on the right side above we have omitted the out-part on $\mathcal{H}^+$ by $\mathcal{CPT}$ invariance and one can easily restore this part by the symmetry.
We choose $T=\frac{1}{z-w}$, and thus 
\begin{align}   \partial_{\bar{z}}T(z,\bar{z})=2\pi\delta^2(z-w)\, .
\end{align}
From this choice, the soft charge \eqref{soft supertranslation charge} is reduced to 
\begin{equation}
    \begin{aligned}
    Q^S    
    &=-\frac{4\pi^2i}{r_s}\lim_{\omega\rightarrow 0^+}\left[\omega\frac{\partial_{\bz}\Tilde{\mathcal{N}}^0_{zz}}{\ga}\right]\ ,
\end{aligned}
\end{equation}
where $\Tilde{\mathcal{N}}_{zz}$ is defined by a Fourier relation $\mathcal{N}_{zz}=\int_{-\infty}^{+\infty}\td u e^{i\omega u} \Tilde{\mathcal{N}}_{zz}^0$ .
After some tedious but direct computations involving the mode expansion \eqref{mode expansion}, one can obtain 
\begin{align}
     Q^S =-16i\frac{r_s^2}{R}\frac{\partial_{\bz}\ga}{\ga}\lim_{\omega\rightarrow0^+}\left[\omega^2a_{+}+\omega^2a_{-}^\dagger\right]\, .
\end{align}
Thus $Q^S$ acts on the in-state as 
\begin{align}\label{soft charge act}
    Q^S\left|{\rm in}\right\rangle=-16i\frac{r_s^2}{R}\frac{\partial_{\bz}\ga}{\ga}\lim_{\omega\rightarrow 0^+}\omega^2a_{-}^\dagger\left|{\rm in}\right\rangle.
\end{align}
As expected, the soft charge creates a low-energy soft graviton in the near horizon state.

Regarding to the hard charge, applying the commutation relation \eqref{commutation hard}, we get 
\begin{align}\label{hard charge act}
   Q^H \left|{\rm in}\right\rangle=\frac{1}{2} \sum_{k=1}^{n}\frac{E_k}{(z-z_k)}\left|{\rm in}\right\rangle\, .
\end{align}
Finally, we can obtain a soft graviton theorem in coordinate space from the Ward identity \eqref{ward identity} from the insertion of \eqref{soft charge act} and \eqref{hard charge act} as
\begin{align}\label{soft theorem coor}
    \partial_{\bz}\left[\ga\lim_{\omega\rightarrow 0^+}a^\dagger_{-}\left|{\rm in}\right\rangle\right]=\lim_{\omega\rightarrow 0^+}\left[\frac{\pi}{32i}\frac{ R}{r_s^2\omega}\ga\sum_{k=1}^n\frac{E_k}{\omega(z-z_k)}\left|{\rm in}\right\rangle\right]\, .
\end{align}

\subsection{Soft graviton theorem in momentum space}

Here we will use the null parametrization in \eqref{graviton momentum} to rewrite the soft theorem derived previously in momentum space. In particular, for each hard particle, the momentum can be parametrized as 
\begin{equation}
\begin{aligned}
q_{k\mu}=&\frac{E_k}{1+z_k\bz_k}\frac{(4\rho+r_s)^3}{16\rho^2(4\rho-r_s)}\\
&\times\left[\frac{16\rho^2(r_s-4\rho)}{(4\rho+r_s)^3}(1+z_k\bz_k),(z_k+\bz_k),-i(z_k-\bz_k),(1-z_k\bz_k)\right]\, .
\end{aligned}
\end{equation}
Applying the null parametrization, one can easily check that
\begin{equation}\label{soft factor momentum}
    \begin{aligned}
    \partial_{\bz}\left(\ga\sum_{k=1}^n\frac{[q_k\cdot \epsilon(p)]^2}{p\cdot q_k}\right)&=-\ga\sum_{k=1}^n\frac{E_k}{\omega(z-z_k)}\left(  1+\frac{\bz_k(z-z_k)}{1+z_k\bz_k}\right)\\
    &=-\ga\sum_{k=1}^n\frac{E_k}{\omega(z-z_k)}\, ,
\end{aligned}
\end{equation}
where we have used the fact that a combination of two components of momenta is also conserved
\begin{align}
   \sum_{k=1}^n \frac{E_k\bz_k}{1+z_k\bz_k}\propto\sum_{k=1}^n \left(q_{k1}-iq_{k2}\right)=0\, .
\end{align}
Finally, inserting \eqref{soft factor momentum}  into \eqref{soft theorem coor}, we arrive at the soft graviton theorem in the momentum space as
\begin{equation}
    \begin{aligned}
    \lim_{\omega\rightarrow 0^+}&\langle {\rm out}|a_{+}\mathcal{S}-\mathcal{S}a^\dagger_{-}\left|{\rm in}\right\rangle\\
    &=\lim_{\omega\rightarrow 0^+}i\frac{\pi R}{32 \omega r_s^2}\left[\sum_{l=1}^m\frac{(q_l^{\rm out}\cdot \epsilon)^2}{p\cdot q_l^{\rm out}}-\sum_{k=1}^n\frac{(q_k^{\rm in}\cdot \epsilon)^2}{p\cdot q_k^{\rm in}}\right]\langle{\rm out}|\mathcal{S}\left|{\rm in}\right\rangle\,,
\end{aligned}
\end{equation}
where we have dropped the derivative from both sides.

\section{Concluding remarks}

To conclude, we study a linearized gravity theory in the near horizon region of the Schwarzschild spacetime. We compute the near horizon symmetry and the on shell surface charges associated to the near horizon symmetries. A soft graviton theorem is obtained from the Ward identity of the near horizon supertranslation. Together with the previous works for gauge theories \cite{Cheng:2022xgm,Cheng:2022xyr}, they constitute a complete realization of the equivalence between the soft theorem and near horizon symmetry in curved spacetime and confirm that the black hole soft hairs are indeed soft particles on the horizon.

To close this paper, we make two remarks about some subtle points. First, it is about the validity of our results at a quantum gravity level in the region of the black hole horizon. The soft theorem derived in this work is based on the assumption that the triangle equivalence at infinity can be extended to the near horizon region of the Schwarzschild black hole. However this extension is not natural at all. The infinity is the region with the weak-field limit, while the near horizon region is at the strong gravity regime. A relevant subtlety is about the soft limit near the black hole horizon, which should involve the mass parameter of the black hole rather than simply comparing the soft and hard external particles \cite{Gaddam:2020mwe,Gaddam:2020rxb}. We do not have a clear answer to those questions. The point of the present computation is that if the triangle relation at the infinity can be extended to the near horizon region, the formulas at the infinity must be able to be repeated in the near horizon region. This is not a easy task a priori. As we have shown in the main text, the split of the soft and hard modes in the near horizon region needs a proper arrangement of the near horizon fall-off conditions. A second subtle point is about the observers. A black hole horizon is a special null hypersurface with respect to observer at infinity. It is nothing special for an infalling observer. Nevertheless, asymptotic analysis indeed has observer dependence, see, for instance the studies in the Rindler space \cite{Chung:2010ge,Takeuchi:2023nxi}, where some different results than the usual flat spacetime were revealed. In general, different gauge and boundary conditions can lead to completely different results for the asymptotic symmetries. So the computation in this work may have no indication at all to an infalling observer. The point is that one needs to initiate a completely different setup for an infalling observer, namely, constructing the coordinate system, imposing proper gauge and boundary conditions in this coordinate system. Then it is able to ask what is the manifestation of near horizon symmetries in the infalling frame.

\section*{Acknowledgments}

The authors thank Peng Cheng for the early collaboration on this project and discussions which helped us clarify some aspects of this work. This work is supported in part by the National Natural Science Foundation of China (NSFC) under Grants No. 11905156 and No. 11935009. K.Y.Z. is also supported in part by NSFC Grant No. 11905158.

\appendix
\section{Modification of stress tensor}
\label{stresstensor}

In flat spacetime, a conserved stress tensor is ambiguous up to a rank four tensor \cite{Conde:2016rom}, which can be used to modify a stress tensor. In the curved spacetime, such ambiguity disappears due to the existence of curvature. Nevertheless, one can modify it by adding a divergence-free symmetric rank two tensor. To our purpose, the modified stress tensor satisfies the gauge condition \eqref{condition}. In this work, we are interested in the stress tensor of a massless complex scalar field which is minimally coupled to gravity, thus 
\begin{align}
    \Tilde{T}_{\mu\nu}\equiv \frac{1}{\sqrt{-g}}\frac{\delta(\sqrt{-g}\mathcal{L}_{M})}{\delta g_{\mu\nu}}=\frac{1}{2}\left(\partial_\mu\Phi\partial_\nu\bar{\Phi}+\partial_\nu\Phi\partial_\mu\bar{\Phi}\right)-\frac{1}{2}g_{\mu\nu}\nabla^\rho\Phi\nabla_\rho\bar{\Phi}\, ,
\end{align}
where   $\mathcal{L}_M=\nabla_\mu\Phi\nabla^\mu\bar{\Phi}\,$.  We assume those scalar fields are given in the form of near horizon expansions as  
\begin{align}
    \Phi=\phi+\sum_{n=1}^{\infty}\lambda^n \Phi^{(n)}\, ,\quad 
\bar{\Phi}=\bar{\phi}+\sum_{n=1}^{\infty}\lambda^n \bar{\Phi}^{(n)}\, .
\end{align}
We construct a stress tensor by $T_{\mu\nu}=\Tilde{T}_{\mu\nu}-\Upsilon_{\mu\nu}$, where $T_{r\mu}=0$ or $\Upsilon_{r\mu}=\Tilde{T}_{r\mu}$, and $\Upsilon_{\mu\nu}$ is an arbitrary divergence-free symmetric 2 tensor. Now we use the divergence-free conditions $\nabla^\mu \Upsilon_{\mu\nu}=0$ to fix it. We expand $\Upsilon_{\mu\nu}$ as 
\begin{align}
\Upsilon_{\mu\nu}=\Upsilon^0_{\mu\nu}+\sum_{n=1}^\infty\lambda^n \Upsilon_{\mu\nu}^{(n)}\, .
\end{align}
Begin by the simplest equation $\nabla^\mu\Upsilon_{\mu r}=0$, which yields
\begin{equation}
    \begin{aligned}
    \gamma^{AB}\Upsilon_{AB}=r_s\xi D^A\Tilde{T}_{rA}&-r_s^2\xi\partial_\lambda\left(\xi^2 \Tilde{T}_{ru}\right)\\
    &+r_s^2\xi\partial_\lambda\left(f\xi^2\Tilde{T}_{rr}\right)+r_s^2\xi^3\partial_\lambda f\Tilde{T}_{rr}/2-r_s^3\xi^3\partial_u\Tilde{T}_{rr}\, .
\end{aligned}
\end{equation}
Thus the trace part of $\Upsilon_{\mu\nu}$ is totally fixed by $\Tilde{T}_{\mu\nu}$.

The transverse equations $\nabla^\mu\Upsilon_{\mu A}=0$ lead to
\begin{align}
    \partial_\lambda\Upsilon_{uA}&=\frac{\partial_\lambda\left(f\xi^2\Tilde{T}_{rA}\right)}{\xi^2}-r_s\partial_u\Tilde{T}_{rA}-2\frac{\Upsilon_{uA}}{\xi}+\frac{D^B\Upsilon_{BA}}{r_s\xi^2}\, .
\end{align}
All information that could be extracted from the above equations is 
\begin{align}
    \Upsilon^{(1)}_{uA}&=\Tilde{T}_{rA}^0-r_sT_{rA}^0-2\Upsilon^0_{uA}+D^B\Upsilon_{BA}^0/r_s\, ,
\end{align}
and, for $n\ge 1$,
\begin{equation}
    \begin{aligned}
   \Upsilon^{(n+1)}_{uA}=&\frac{1}{n+1}\left\{-r_s\partial_u\Tilde{T}_{rA}^{(n)}+\sum_{m=0}^{n}(-1)^m\Tilde{T}_{rA}^{(n-m)}\right.\\
   &+\sum_{m=0}^{n-1}(-1)^m\left[(1+m)\Tilde{T}_{rA}^{(n-1-m)}+(n-m)\Tilde{T}_{rA}^{(n-m)}\right]\\
   &+\left.\sum_{m=0}^{n}(-1)^m\left[(1+m)D^B\Upsilon_{BA}^{(n-m)}/r_s-2\Upsilon_{uA}^{(n-m)}\right]\right\}\, .
\end{aligned}
\end{equation}
While the leading order $\Upsilon^{0}_{uA}$ is free. We continue with the equation of the $u$ component, which gives
\begin{align}
    \partial_\lambda\Upsilon_{uu}=\frac{\partial_\lambda(f\xi^2\Tilde{T}_{ru})}{\xi^2}-r_s\partial_u\Tilde{T}_{ru}-2\frac{\Upsilon_{uu}}{\xi}+\frac{D^A\Upsilon_{uA}}{r_s\xi^2}\, .
\end{align}
This equation controls $\Upsilon_{uu}$ up to an initial data $\Upsilon^0_{uu}$,
\begin{align}
    \Upsilon^{(1)}_{uu}&=\Tilde{T}_{ru}^0-r_sT_{ru}^0-2\Upsilon^0_{uu}+D^A\Upsilon_{uA}^0/r_s\, ,
\end{align}
and, for $n\ge 1$,
\begin{equation}
    \begin{aligned}
   \Upsilon^{(n+1)}_{uu}=&\frac{1}{n+1}\left\{-r_s\partial_u\Tilde{T}_{ru}^{(n)}+\sum_{m=0}^{n}(-1)^m\Tilde{T}_{ru}^{(n-m)}\right.\\
   &+\sum_{m=0}^{n-1}(-1)^m\left[(1+m)\Tilde{T}_{ru}^{(n-1-m)}+(n-m)\Tilde{T}_{ru}^{(n-m)}\right]\\
   &+\left.\sum_{m=0}^{n}(-1)^m\left[(1+m)D^A\Upsilon_{uA}^{(n-m)}/r_s-2\Upsilon_{uu}^{(n-m)}\right]\right\}\, .
\end{aligned}
\end{equation}
Finally, we find the traceless part of $\Upsilon_{AB}$ and the leading order of $\Upsilon_{uu}$, $\Upsilon_{uA}$ are free. The $uu, uA, zz$ components of the modified stress tensor at the leading orders are
\begin{align}
    T_{uu}^{0}&=\partial_u\phi\partial_u\bar{\phi}-\Upsilon_{uu}^0\, ,\\
    T_{uA}^{0}&=\frac{1}{2}\left(\partial_u\phi\partial_A\bar{\phi}+\partial_A\phi\partial_u\bar{\phi}\right)-\Upsilon_{uA}^0\, ,\\
    T_{zz}^0&=\partial_z\phi\partial_z\bar{\phi}-\Upsilon_{zz}^0\, .
\end{align}
The hard part of the near horizon supertranslation charge \eqref{hard} in the main text is sensitive to the modification of the stress tensor. In particular, different choices of the leading orders of $\Upsilon_{uu}$, $\Upsilon_{uA}$ will yield different formulas of soft theorem. In this work, we set $\Upsilon_{uA}^0=0$, and use the free initial data $\Upsilon_{uA}^0$ to set $T_{uA}^{0}=0$.

\bibliography{ref}

\providecommand{\href}[2]{#2}\begingroup\raggedright\begin{thebibliography}{10}

\bibitem{Bardeen:1973gs}
J.~M. Bardeen, B.~Carter, and S.~W. Hawking, ``{The Four laws of black hole
  mechanics},'' \href{http://dx.doi.org/10.1007/BF01645742}{{\em Commun. Math.
  Phys.} {\bfseries 31} (1973) 161--170}.

\bibitem{Hawking:1975vcx}
S.~W. Hawking, ``{Particle Creation by Black Holes},''
  \href{http://dx.doi.org/10.1007/BF02345020}{{\em Commun. Math. Phys.}
  {\bfseries 43} (1975) 199--220}. [Erratum: Commun.Math.Phys. 46, 206 (1976)].

\bibitem{Hawking:1976ra}
S.~W. Hawking, ``{Breakdown of Predictability in Gravitational Collapse},''
  \href{http://dx.doi.org/10.1103/PhysRevD.14.2460}{{\em Phys. Rev. D}
  {\bfseries 14} (1976) 2460--2473}.

\bibitem{Hawking:2016msc}
S.~W. Hawking, M.~J. Perry, and A.~Strominger, ``{Soft Hair on Black Holes},''
  \href{http://dx.doi.org/10.1103/PhysRevLett.116.231301}{{\em Phys. Rev.
  Lett.} {\bfseries 116} no.~23, (2016) 231301},
  \href{http://arxiv.org/abs/1601.00921}{{\ttfamily arXiv:1601.00921
  [hep-th]}}.

\bibitem{Strominger:2017zoo}
A.~Strominger, {\em {Lectures on the Infrared Structure of Gravity and Gauge
  Theory}}.
\newblock Princeton University Press, Princeton, 2018.
\newblock \href{http://arxiv.org/abs/1703.05448}{{\ttfamily arXiv:1703.05448
  [hep-th]}}.

\bibitem{Strominger:2013lka}
A.~Strominger, ``{Asymptotic Symmetries of Yang-Mills Theory},''
  \href{http://dx.doi.org/10.1007/JHEP07(2014)151}{{\em JHEP} {\bfseries 07}
  (2014) 151}, \href{http://arxiv.org/abs/1308.0589}{{\ttfamily arXiv:1308.0589
  [hep-th]}}.

\bibitem{Strominger:2013jfa}
A.~Strominger, ``{On BMS Invariance of Gravitational Scattering},''
  \href{http://dx.doi.org/10.1007/JHEP07(2014)152}{{\em JHEP} {\bfseries 07}
  (2014) 152}, \href{http://arxiv.org/abs/1312.2229}{{\ttfamily arXiv:1312.2229
  [hep-th]}}.

\bibitem{He:2014laa}
T.~He, V.~Lysov, P.~Mitra, and A.~Strominger, ``{BMS supertranslations and
  Weinberg\textquoteright{}s soft graviton theorem},''
  \href{http://dx.doi.org/10.1007/JHEP05(2015)151}{{\em JHEP} {\bfseries 05}
  (2015) 151}, \href{http://arxiv.org/abs/1401.7026}{{\ttfamily arXiv:1401.7026
  [hep-th]}}.

\bibitem{He:2014cra}
T.~He, P.~Mitra, A.~P. Porfyriadis, and A.~Strominger, ``{New Symmetries of
  Massless QED},'' \href{http://dx.doi.org/10.1007/JHEP10(2014)112}{{\em JHEP}
  {\bfseries 10} (2014) 112}, \href{http://arxiv.org/abs/1407.3789}{{\ttfamily
  arXiv:1407.3789 [hep-th]}}.

\bibitem{He:2015zea}
T.~He, P.~Mitra, and A.~Strominger, ``{2D Kac-Moody Symmetry of 4D Yang-Mills
  Theory},'' \href{http://dx.doi.org/10.1007/JHEP10(2016)137}{{\em JHEP}
  {\bfseries 10} (2016) 137}, \href{http://arxiv.org/abs/1503.02663}{{\ttfamily
  arXiv:1503.02663 [hep-th]}}.

\bibitem{Compere:2016jwb}
G.~Comp\`ere and J.~Long, ``{Vacua of the gravitational field},''
  \href{http://dx.doi.org/10.1007/JHEP07(2016)137}{{\em JHEP} {\bfseries 07}
  (2016) 137}, \href{http://arxiv.org/abs/1601.04958}{{\ttfamily
  arXiv:1601.04958 [hep-th]}}.

\bibitem{Compere:2016hzt}
G.~Comp\`ere and J.~Long, ``{Classical static final state of collapse with
  supertranslation memory},''
  \href{http://dx.doi.org/10.1088/0264-9381/33/19/195001}{{\em Class. Quant.
  Grav.} {\bfseries 33} no.~19, (2016) 195001},
  \href{http://arxiv.org/abs/1602.05197}{{\ttfamily arXiv:1602.05197 [gr-qc]}}.

\bibitem{Mao:2016pwq}
P.~Mao, X.~Wu, and H.~Zhang, ``{Soft hairs on isolated horizon implanted by
  electromagnetic fields},''
  \href{http://dx.doi.org/10.1088/1361-6382/aa59da}{{\em Class. Quant. Grav.}
  {\bfseries 34} no.~5, (2017) 055003},
  \href{http://arxiv.org/abs/1606.03226}{{\ttfamily arXiv:1606.03226
  [hep-th]}}.

\bibitem{Afshar:2016uax}
H.~Afshar, D.~Grumiller, and M.~M. Sheikh-Jabbari, ``{Near horizon soft hair as
  microstates of three dimensional black holes},''
  \href{http://dx.doi.org/10.1103/PhysRevD.96.084032}{{\em Phys. Rev. D}
  {\bfseries 96} no.~8, (2017) 084032},
  \href{http://arxiv.org/abs/1607.00009}{{\ttfamily arXiv:1607.00009
  [hep-th]}}.

\bibitem{Mirbabayi:2016axw}
M.~Mirbabayi and M.~Porrati, ``{Dressed Hard States and Black Hole Soft
  Hair},'' \href{http://dx.doi.org/10.1103/PhysRevLett.117.211301}{{\em Phys.
  Rev. Lett.} {\bfseries 117} no.~21, (2016) 211301},
  \href{http://arxiv.org/abs/1607.03120}{{\ttfamily arXiv:1607.03120
  [hep-th]}}.

\bibitem{Grumiller:2016kcp}
D.~Grumiller, A.~Perez, S.~Prohazka, D.~Tempo, and R.~Troncoso, ``{Higher Spin
  Black Holes with Soft Hair},''
  \href{http://dx.doi.org/10.1007/JHEP10(2016)119}{{\em JHEP} {\bfseries 10}
  (2016) 119}, \href{http://arxiv.org/abs/1607.05360}{{\ttfamily
  arXiv:1607.05360 [hep-th]}}.

\bibitem{Donnay:2016ejv}
L.~Donnay, G.~Giribet, H.~A. Gonz\'alez, and M.~Pino, ``{Extended Symmetries at
  the Black Hole Horizon},''
  \href{http://dx.doi.org/10.1007/JHEP09(2016)100}{{\em JHEP} {\bfseries 09}
  (2016) 100}, \href{http://arxiv.org/abs/1607.05703}{{\ttfamily
  arXiv:1607.05703 [hep-th]}}.

\bibitem{Cai:2016idg}
R.-G. Cai, S.-M. Ruan, and Y.-L. Zhang, ``{Horizon supertranslation and
  degenerate black hole solutions},''
  \href{http://dx.doi.org/10.1007/JHEP09(2016)163}{{\em JHEP} {\bfseries 09}
  (2016) 163}, \href{http://arxiv.org/abs/1609.01056}{{\ttfamily
  arXiv:1609.01056 [gr-qc]}}.

\bibitem{Hawking:2016sgy}
S.~W. Hawking, M.~J. Perry, and A.~Strominger, ``{Superrotation Charge and
  Supertranslation Hair on Black Holes},''
  \href{http://dx.doi.org/10.1007/JHEP05(2017)161}{{\em JHEP} {\bfseries 05}
  (2017) 161}, \href{http://arxiv.org/abs/1611.09175}{{\ttfamily
  arXiv:1611.09175 [hep-th]}}.

\bibitem{Afshar:2016kjj}
H.~Afshar, D.~Grumiller, W.~Merbis, A.~Perez, D.~Tempo, and R.~Troncoso,
  ``{Soft hairy horizons in three spacetime dimensions},''
  \href{http://dx.doi.org/10.1103/PhysRevD.95.106005}{{\em Phys. Rev. D}
  {\bfseries 95} no.~10, (2017) 106005},
  \href{http://arxiv.org/abs/1611.09783}{{\ttfamily arXiv:1611.09783
  [hep-th]}}.

\bibitem{Choi:2018oel}
S.~Choi and R.~Akhoury, ``{Soft Photon Hair on Schwarzschild Horizon from a
  Wilson Line Perspective},''
  \href{http://dx.doi.org/10.1007/JHEP12(2018)074}{{\em JHEP} {\bfseries 12}
  (2018) 074}, \href{http://arxiv.org/abs/1809.03467}{{\ttfamily
  arXiv:1809.03467 [hep-th]}}.

\bibitem{Choi:2019fuq}
S.~Choi, S.~Sandeep~Pradhan, and R.~Akhoury, ``{Supertranslation Hair of
  Schwarzschild Black Hole: A Wilson Line Perspective},''
  \href{http://dx.doi.org/10.1007/JHEP01(2020)013}{{\em JHEP} {\bfseries 01}
  (2020) 013}, \href{http://arxiv.org/abs/1910.05882}{{\ttfamily
  arXiv:1910.05882 [hep-th]}}.

\bibitem{Cheng:2022xyr}
P.~Cheng and P.~Mao, ``{Soft theorems in curved spacetime},''
  \href{http://dx.doi.org/10.1103/PhysRevD.106.L081702}{{\em Phys. Rev. D}
  {\bfseries 106} no.~8, (2022) L081702},
  \href{http://arxiv.org/abs/2206.11564}{{\ttfamily arXiv:2206.11564
  [hep-th]}}.

\bibitem{Cheng:2022xgm}
P.~Cheng and P.~Mao, ``{Soft gluon theorems in curved spacetime},''
  \href{http://dx.doi.org/10.1103/PhysRevD.107.065010}{{\em Phys. Rev. D}
  {\bfseries 107} no.~6, (2023) 065010},
  \href{http://arxiv.org/abs/2211.00031}{{\ttfamily arXiv:2211.00031
  [hep-th]}}.

\bibitem{Mao:2023rca}
P.~Mao and K.-Y. Zhang, ``{Soft theorems in de Sitter spacetime},''
  \href{http://arxiv.org/abs/2308.08861}{{\ttfamily arXiv:2308.08861
  [hep-th]}}.

\bibitem{Newman:1962cia}
E.~T. Newman and T.~W.~J. Unti, ``{Behavior of Asymptotically Flat Empty
  Spaces},'' \href{http://dx.doi.org/10.1063/1.1724303}{{\em J. Math. Phys.}
  {\bfseries 3} no.~5, (1962) 891}.

\bibitem{Adami:2021nnf}
H.~Adami, D.~Grumiller, M.~M. Sheikh-Jabbari, V.~Taghiloo, H.~Yavartanoo, and
  C.~Zwikel, ``{Null boundary phase space: slicings, news \& memory},''
  \href{http://dx.doi.org/10.1007/JHEP11(2021)155}{{\em JHEP} {\bfseries 11}
  (2021) 155}, \href{http://arxiv.org/abs/2110.04218}{{\ttfamily
  arXiv:2110.04218 [hep-th]}}.

\bibitem{Donnay:2015abr}
L.~Donnay, G.~Giribet, H.~A. Gonzalez, and M.~Pino, ``{Supertranslations and
  Superrotations at the Black Hole Horizon},''
  \href{http://dx.doi.org/10.1103/PhysRevLett.116.091101}{{\em Phys. Rev.
  Lett.} {\bfseries 116} no.~9, (2016) 091101},
  \href{http://arxiv.org/abs/1511.08687}{{\ttfamily arXiv:1511.08687
  [hep-th]}}.

\bibitem{Adami:2020amw}
H.~Adami, D.~Grumiller, S.~Sadeghian, M.~M. Sheikh-Jabbari, and C.~Zwikel,
  ``{T-Witts from the horizon},''
  \href{http://dx.doi.org/10.1007/JHEP04(2020)128}{{\em JHEP} {\bfseries 04}
  (2020) 128}, \href{http://arxiv.org/abs/2002.08346}{{\ttfamily
  arXiv:2002.08346 [hep-th]}}.

\bibitem{Barnich:2004ts}
G.~Barnich, S.~Leclercq, and P.~Spindel, ``{Classification of surface charges
  for a spin two field on a curved background solution},''
  \href{http://dx.doi.org/10.1023/B:MATH.0000045554.71211.99}{{\em Lett. Math.
  Phys.} {\bfseries 68} (2004) 175--181},
  \href{http://arxiv.org/abs/gr-qc/0404006}{{\ttfamily arXiv:gr-qc/0404006}}.

\bibitem{Conde:2016rom}
E.~Conde and P.~Mao, ``{BMS Supertranslations and Not So Soft Gravitons},''
  \href{http://dx.doi.org/10.1007/JHEP05(2017)060}{{\em JHEP} {\bfseries 05}
  (2017) 060}, \href{http://arxiv.org/abs/1612.08294}{{\ttfamily
  arXiv:1612.08294 [hep-th]}}.

\bibitem{Bondi:1962px}
H.~Bondi, M.~G.~J. van~der Burg, and A.~W.~K. Metzner, ``{Gravitational waves
  in general relativity. 7. Waves from axisymmetric isolated systems},''
  \href{http://dx.doi.org/10.1098/rspa.1962.0161}{{\em Proc. Roy. Soc. Lond. A}
  {\bfseries 269} (1962) 21--52}.

\bibitem{Sachs:1962wk}
R.~K. Sachs, ``{Gravitational waves in general relativity. 8. Waves in
  asymptotically flat space-times},''
  \href{http://dx.doi.org/10.1098/rspa.1962.0206}{{\em Proc. Roy. Soc. Lond. A}
  {\bfseries 270} (1962) 103--126}.

\bibitem{Barnich:2010eb}
G.~Barnich and C.~Troessaert, ``{Aspects of the BMS/CFT correspondence},''
  \href{http://dx.doi.org/10.1007/JHEP05(2010)062}{{\em JHEP} {\bfseries 05}
  (2010) 062}, \href{http://arxiv.org/abs/1001.1541}{{\ttfamily arXiv:1001.1541
  [hep-th]}}.

\bibitem{Aggarwal:2023qwl}
A.~Aggarwal and N.~Gaddam, ``{All symmetries of near-horizon scattering},''
  \href{http://arxiv.org/abs/2309.05775}{{\ttfamily arXiv:2309.05775
  [hep-th]}}.

\bibitem{Barnich:2001jy}
G.~Barnich and F.~Brandt, ``{Covariant theory of asymptotic symmetries,
  conservation laws and central charges},''
  \href{http://dx.doi.org/10.1016/S0550-3213(02)00251-1}{{\em Nucl. Phys. B}
  {\bfseries 633} (2002) 3--82},
  \href{http://arxiv.org/abs/hep-th/0111246}{{\ttfamily arXiv:hep-th/0111246}}.

\bibitem{Adami:2021kvx}
H.~Adami, M.~M. Sheikh-Jabbari, V.~Taghiloo, and H.~Yavartanoo, ``{Null surface
  thermodynamics},'' \href{http://dx.doi.org/10.1103/PhysRevD.105.066004}{{\em
  Phys. Rev. D} {\bfseries 105} no.~6, (2022) 066004},
  \href{http://arxiv.org/abs/2110.04224}{{\ttfamily arXiv:2110.04224
  [hep-th]}}.

\bibitem{Adami:2023fbm}
H.~Adami, A.~Parvizi, M.~M. Sheikh-Jabbari, V.~Taghiloo, and H.~Yavartanoo,
  ``{Hydro \& thermo dynamics at causal boundaries, examples in 3d gravity},''
  \href{http://dx.doi.org/10.1007/JHEP07(2023)038}{{\em JHEP} {\bfseries 07}
  (2023) 038}, \href{http://arxiv.org/abs/2305.01009}{{\ttfamily
  arXiv:2305.01009 [hep-th]}}.

\bibitem{Adami:2020ugu}
H.~Adami, M.~M. Sheikh-Jabbari, V.~Taghiloo, H.~Yavartanoo, and C.~Zwikel,
  ``{Symmetries at null boundaries: two and three dimensional gravity cases},''
  \href{http://dx.doi.org/10.1007/JHEP10(2020)107}{{\em JHEP} {\bfseries 10}
  (2020) 107}, \href{http://arxiv.org/abs/2007.12759}{{\ttfamily
  arXiv:2007.12759 [hep-th]}}.

\bibitem{Ruzziconi:2020wrb}
R.~Ruzziconi and C.~Zwikel, ``{Conservation and Integrability in
  Lower-Dimensional Gravity},''
  \href{http://dx.doi.org/10.1007/JHEP04(2021)034}{{\em JHEP} {\bfseries 04}
  (2021) 034}, \href{http://arxiv.org/abs/2012.03961}{{\ttfamily
  arXiv:2012.03961 [hep-th]}}.

\bibitem{Adami:2021sko}
H.~Adami, M.~M. Sheikh-Jabbari, V.~Taghiloo, H.~Yavartanoo, and C.~Zwikel,
  ``{Chiral Massive News: Null Boundary Symmetries in Topologically Massive
  Gravity},'' \href{http://dx.doi.org/10.1007/JHEP05(2021)261}{{\em JHEP}
  {\bfseries 05} (2021) 261}, \href{http://arxiv.org/abs/2104.03992}{{\ttfamily
  arXiv:2104.03992 [hep-th]}}.

\bibitem{Elvang:2016qvq}
H.~Elvang, C.~R.~T. Jones, and S.~G. Naculich, ``{Soft Photon and Graviton
  Theorems in Effective Field Theory},''
  \href{http://dx.doi.org/10.1103/PhysRevLett.118.231601}{{\em Phys. Rev.
  Lett.} {\bfseries 118} no.~23, (2017) 231601},
  \href{http://arxiv.org/abs/1611.07534}{{\ttfamily arXiv:1611.07534
  [hep-th]}}.

\bibitem{Lee:1990nz}
J.~Lee and R.~M. Wald, ``{Local symmetries and constraints},''
  \href{http://dx.doi.org/10.1063/1.528801}{{\em J. Math. Phys.} {\bfseries 31}
  (1990) 725--743}.

\bibitem{Iyer:1994ys}
V.~Iyer and R.~M. Wald, ``{Some properties of Noether charge and a proposal for
  dynamical black hole entropy},''
  \href{http://dx.doi.org/10.1103/PhysRevD.50.846}{{\em Phys. Rev. D}
  {\bfseries 50} (1994) 846--864},
  \href{http://arxiv.org/abs/gr-qc/9403028}{{\ttfamily arXiv:gr-qc/9403028}}.

\bibitem{Gaddam:2020mwe}
N.~Gaddam and N.~Groenenboom, ``{Soft graviton exchange and the information
  paradox},'' \href{http://dx.doi.org/10.1103/PhysRevD.109.026007}{{\em Phys.
  Rev. D} {\bfseries 109} no.~2, (2024) 026007},
  \href{http://arxiv.org/abs/2012.02355}{{\ttfamily arXiv:2012.02355
  [hep-th]}}.

\bibitem{Gaddam:2020rxb}
N.~Gaddam, N.~Groenenboom, and G.~'t~Hooft, ``{Quantum gravity on the black
  hole horizon},'' \href{http://dx.doi.org/10.1007/JHEP01(2022)023}{{\em JHEP}
  {\bfseries 01} (2022) 023}, \href{http://arxiv.org/abs/2012.02357}{{\ttfamily
  arXiv:2012.02357 [hep-th]}}.

\bibitem{Chung:2010ge}
H.~Chung, ``{Asymptotic Symmetries of Rindler Space at the Horizon and Null
  Infinity},'' \href{http://dx.doi.org/10.1103/PhysRevD.82.044019}{{\em Phys.
  Rev. D} {\bfseries 82} (2010) 044019},
  \href{http://arxiv.org/abs/1005.0820}{{\ttfamily arXiv:1005.0820 [gr-qc]}}.

\bibitem{Takeuchi:2023nxi}
S.~Takeuchi, ``{Asymptotic Gauge Symmetry in Rindler Coordinates with Canonical
  Quantized U(1) Gauge Field},''
  \href{http://arxiv.org/abs/2309.09798}{{\ttfamily arXiv:2309.09798
  [hep-th]}}.

\end{thebibliography}\endgroup

\end{document}